\documentclass[twocolumn]{revtex4}
\usepackage{txfonts}
\usepackage{graphicx}
\usepackage{dcolumn}
\usepackage{bm}
\usepackage[usenames]{color}

\begin{document}
\title{
On-chip quantum feedback control of a superconducting qubit 
}
\author{K. Kakuyanagi,$^{1}$ A. Kemp,$^{1}$ T. Baba,$^{1,2}$ Y. Matsuzaki,$^{1}$ H. Nakano,$^{1}$ 
K. Semba,$^{1,3}$  S. Saito,$^{1}$}
\affiliation{
$^{1}$NTT Basic Research Laboratories, NTT Corporation, Kanagawa, 243-0198, Japan\\
$^{2}$Tokyo University of Science, 1-3 Kagurazaka, Shinjuku, Tokyo 162-8601, Japan\\
$^{3}$National Institute of Information and Communications Technology, Koganei, Tokyo, 184-8795, Japan.\\
}

\begin{abstract}
Quantum feedback is a technique for measuring a qubit and applying appropriate feedback depending on the measurement results.
Here, we propose a new on-chip quantum feedback method where the measurement-result information is not taken from the chip to the outside of a dilution refrigerator. 
 This can be done by using a selective qubit-energy shift induced by measurement apparatus.
We demonstrate on-chip quantum feedback and succeed in the rapid initialization of a qubit by flipping the qubit state only when we detect the ground state of the qubit. 
The feedback loop of our quantum feedback method closed on a chip, and so the operating time needed to control a qubit is of the order of 10 ns. 
This operating time is shorter than with the convectional off-chip feedback method. 
Our on-chip quantum feedback technique opens many possibilities such as an application to quantum information processing and providing an understanding of the foundation of thermodynamics for quantum systems.
\end{abstract}

\maketitle

Quantum feedback (or feed-forward) is a technique for controlling a qubit state depending on the measurement result of the qubit \cite{QFB1,QFB2,QFB3,QFB4,QFB5,QFB6,QFB7,QFB8}.
Quantum measurements change the quantum states depending on measurement results, and such a projection process is probabilistic. 
Interestingly, quantum feedback enables us to make the process deterministic due to the measurement-result-dependent control on the system.
In particular, quantum feedback is essential for quantum information processing such as quantum-state stabilization \cite{QFB9}, quantum control \cite{quantumcontrol1,quantumcontrol2}, initialization\cite{initialization}, entanglement generation \cite{entanglementgeneration}, one-way quantum computation \cite{oneway}, and quantum error corrections \cite{QEC}.
Also, quantum feedback is known to be related to the concept of Maxwell's demon \cite{demon1, demon2}. 
When noise information is given by quantum measurements, quantum feedback enables us to suppress quantum, thermal, and technical noise \cite{demon3}. 
Moreover, it is in principle possible to perform conversion from information to energy via a feedback protocol \cite{demon4}. 
So the quantum feedback technique
is essential to realize quantum information processing and to understand the foundation of thermodynamics.

A superconducting qubit is one candidate for realizing scalable quantum information processing \cite{SCQ1,SCQ2,SCQ2b,SCQ3,SCQ4}.
Also, a superconducting qubit could in principle provide thermodynamic gain via quantum feedback control, and several theoretical proposals have been made \cite{demon5,demon6,demon7}.
However, because of a strong interaction with the environment, the superconducting qubit has a relatively short coherence time compared with other qubits such as an electron spin or a trapped atom.
So, if we use superconducting qubits to realize quantum information processing or to obtain significant thermodynamic gain, we should perform quantum feedback much faster than the short coherence time of a superconducting qubit.

The simple way to realize quantum feedback is to generate and apply qubit control pulses after obtaining information about the measured qubit state \cite{conventional,conventional2}.
However, in this conventional method, there is a significant time delay due to an off-chip information flow including data processing.
For example, we need time for the measurement result to reach the signal processor we use, time to generate control pulses depending on the result (a few microseconds), and  time for the control pulses to reach the qubit.
It is worth mentioning that, with a superconducting qubit, a qubit chip is prepared in a dilution refrigerator, and measurement equipment is set at room temperature.
So, we have to connect a qubit chip to the measurement equipment using long cables.
The feedback loop of this measurement system is at least around 10 m.
If the cable delay time is 5 ns/m, the delay time on the cable is about 100 ns for a 20 m feedback loop. 
Such a conventional feedback process might be too slow to realize quantum information processing or to obtain significant thermodynamic gain especially when many quantum feedbacks are required within the coherence time of the superconducting qubit.
The best way to suppress the time delay is an on-chip information flow as we describe later.

To read out a superconducting flux qubit, we use a Josephson bifurcation amplifier (JBA) technique \cite{JBA}.
The JBA readout method uses a nonlinear resonator coupled to the flux qubit.
Because of the nonlinearity, the resonator of the JBA exhibits a bistable state when it is driven at the optimum power and frequency.
By using a microwave readout pulse to drive a JBA resonator, we can perform a projective measurement of the qubit. 
The coupling between a qubit and a JBA resonator
shows that the JBA resonator exhibits a different time evolution depending on the state of the qubit, and so the readout of the state of the JBA resonator gives us information about the qubit state.
Moreover, it is known that the JBA readout method preserves the expected value of the measured observable reading the qubit before and after the measurement, and so this provides us with the ability to implement quantum non-demolition (QND) measurements \cite{QND1}.
Low projection error is one of the characteristics of the JBA readout method \cite{pj_err}. 
This means that, when we prepare the qubit excited (ground) state, the JBA resonator accurately projects the qubit to the excited (ground) state.

\begin{figure}[h]
\begin{center}
\includegraphics[width=1.0\columnwidth]{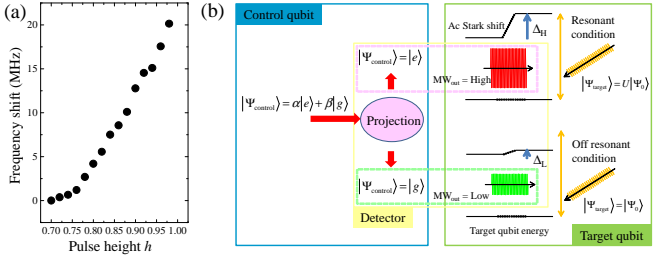}
\caption{
(a)	
A height dependence of the qubit-energy shift.
This shows the observed qubit energy shift in our previous measurement \cite{OURS}.
As we increase the height of an applied readout pulse, the frequency shift also increases.
(b) Principle of on-chip quantum feedback.
Due to the AC Stark shift, we can rotate the target qubit only when we project the control-qubit into an excited state, while no rotation occurs when projecting the control-qubit into a ground state.
}
\label{feedback}
\end{center}
\vspace{-0.2cm}
\end{figure}

In our previous measurement, we found that the qubit energy shifted depending on the height of the readout pulse \cite{OURS}.
Fig. \ref{feedback}(a) shows the readout-pulse height dependence of the qubit energy shift in the measurement.
The qubit energy was shifted due to the coupling between the qubit and the readout microwave pulse.
This phenomenon is considered to be the same as the AC Stark shift \cite{acs1,acs2}. 
In this letter, as the first step of complete control of the qubit state, we demonstrate ultra-fast single qubit initialization by the on-chip quantum feedback technique where we utilize the AC Stark shift of the qubit.

Fig. \ref{feedback}(b) shows our proposed general on-chip quantum feedback technique.
Firstly, we perform a measurement to readout a control-qubit state using a JBA readout method.
The control qubit is projected into an excited or a ground state by the measurement.
The bistable states of the JBA resonance show high or low amplitudes.
After the bifurcation process, the output of the transmission microwave amplitude depends on the state of the JBA, and the JBA resonator exhibits a high (low) amplitude state when we detect a control qubit in an excited (ground) state.
There is electric magnetic coupling between the target qubit and the output microwave, which induces an energy shift of the target qubit due to the AC Stark effect.
The amount of the energy shift depends on the amplitude of the coupled microwave. 
Actually, the target qubit energy is $\hbar \omega_H=\hbar\omega_0+\Delta_{H}$ and $\hbar \omega_L=\hbar\omega_0+\Delta_{L}$ for an excited state and a ground state, respectively.
Secondly, we apply a microwave pulse with a frequency $\omega_H$ to the target qubit.
When we detect the excited state of the control qubit via the JBA measurement, the microwave pulse is on resonant and can rotate the target qubit.
On the other hand, when we detect the ground state of the control qubit, it is impossible to rotate the target qubit by the microwave pulse because of the detuning.
Importantly, by using this qubit energy shift when applying a readout pulse, we realize arbitrary control of the target-qubit state depending on the measurement result.
Finally, we turn off the JBA readout pulse.
Although this example shows finite $\Delta_{L}$, we can effectively choose $\Delta_{L}=0$ by using an external magnetic field shift pulse for the target qubit with the microwave pulse.
In this case, the phase modification of the target qubit can be negligible when we detect the ground state of the control qubit.
It is worth mentioning that, since we close the feedback information loop in a small chip, we can avoid the time loss caused by transmitting and receiving the measurement information between the chip and outside of the refrigerator.
Compared with a conventional off-chip method that suffers from such an intrinsic time loss, our on-chip quantum feedback method has the advantage of operational speed.

\begin{figure}[h]
\begin{center}
\includegraphics[width=1.0\columnwidth]{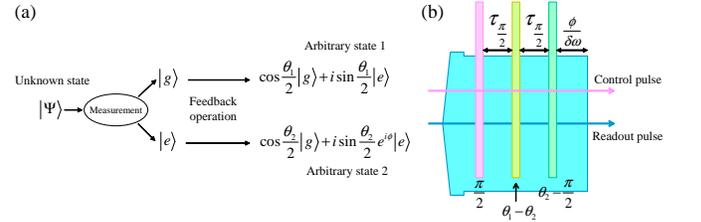}
\caption{
(a) Schematic of a single-qubit quantum feedback.
(b) Pulse sequence for quantum feedback. Three rectangular pulses are composed of qubit-control microwave pulses with a frequency tuned to the qubit energy. 
The pulse distance is determined by a qubit energy shift induced by a readout pulse. The blue pulse denotes a JBA readout pulse with its frequency tuned to a JBA resonator.
}
\label{sequence1}
\end{center}
\vspace{-0.2cm}
\end{figure}

As a concrete example, we describe the quantum feedback control for a single qubit in detail.
Fig. \ref{sequence1}(a) is a schematic image of a single-qubit quantum feedback .
Firstly, an unknown qubit state is projected to a ground or an excited state by the measurement.
Secondly, we perform a different gate operation depending on the measurement result.
As a result, we can deterministically prepare an arbitrary quantum state from the unknown state.
For this purpose, we use the pulse sequence shown in Fig. \ref{sequence1}(b) to demonstrate on-chip single-qubit quantum feedback.
We apply three qubit control pulses during applying the JBA readout pulse. 
The JBA readout pulse probabilistically projects a qubit state to an excited or ground state. Since the readout of the qubit via the JBA is a QND measurement, the qubit remains in the post measurement state after the projection. 
During the application of the readout pulse, the qubit energy has a finite shift whose amplitude depends on the qubit state projected by the JBA.

The Hamiltonian of the qubit and applied microwave are described as follows.
 \begin{eqnarray}
  H={\hbar\omega_{\rm qubit} \over 2}\sigma_z+\hbar\Omega\sigma_x \cos{\omega t}
 \end{eqnarray}
where $\omega_{\rm qubit}$ denotes the qubit resonant frequency, and $\omega$ denotes the applied microwave frequency.
On a rotating frame, by using a rotating wave approximation, the effective Hamiltonian is described as follows.
\begin{equation}
H'={\hbar\over 2}\left(\omega_{\rm qubit}-\omega\right)\sigma_z+{\hbar\Omega \over 2}\sigma_x
\label{hamiltonian}
\end{equation}
The Rabi oscillation frequency depends on the applied microwave power.
In this sequence, the Rabi frequency $\Omega$ is much larger than the qubit frequency difference $\Delta\omega\equiv \omega_{\rm qubit}-\omega$.
Under this condition $(\Omega \gg \Delta\omega)$, we can ignore the first term of equation (\ref{hamiltonian}) when we apply a resonant microwave for Rabi oscillation.
With Ramsey oscillation, the second term of equation (\ref{hamiltonian}) is equal to 0.
So, we can describe the time evolution of the qubit by using two unitary operators $R$ and $T$.
These unitary operators are defined as follows. 
  \begin{eqnarray}
   R\left(\theta\right)=e^{i {\theta \over 2}\sigma_x}\\
   T\left(\tau, \Delta\omega \right)=e^{i {\Delta\omega\tau \over
   2}\sigma_z}
  \end{eqnarray}
where $R$ ($T$) represents the rotation of the $x$ ($z$) axis in a Bloch sphere, which corresponds to Rabi (Ramsey) oscillations.

The first qubit-control pulse rotated a qubit state by an angle of $\pi\over 2$.
When a JBA detects a qubit excited state, the JBA shows a high amplitude state. Compared with a low amplitude state,
 we have a qubit energy shift of $\hbar\delta\omega\equiv\hbar\left(\omega_H-\omega_L\right)$ for the high amplitude case.
Owing to this energy shift, Ramsey rotation occurs only if JBA detects an excited state of the qubit, which we call a selective Ramsey oscillation.
When the qubit is rotated by an angle of $\frac{\pi }{2}$ due to such a selective Ramsey oscillation, we apply the second qubit-control pulse to induce the Rabi oscillation. 
This second pulse rotates the qubit state with an angle of $\left(\theta_1-\theta_2\right)$. 
After waiting for another selective Ramsey ${\pi\over 2}$ oscillation, we apply the third qubit-control pulse. 
The third control pulse rotates a qubit state by an angle of $\left(\theta_2-{\pi\over 2}\right)$. 
Finally, we wait for a time of ${\phi\over \delta \omega}$ to induce the selective Ramsey oscillation, and then we turn off the readout pulse.
The final state realized by this sequence is described as follows.
\begin {eqnarray}
&&\left|\Psi_{\rm final}^{(g)}\right>\nonumber \\
 &=&\! T\!\left(\!{\phi \over
 \delta\omega},\!0\right)\!R\!\left(\!\theta_2\!-\!{\pi\over
 2}\!\right)\!T\!\left(\!\tau_{\pi\over
 2},\!0\!\right)\!R\!\left(\!\theta_1\!-\!\theta_2\!\right)\!T\!\left(\!\tau_{\pi\over
 2},\!0\right)\!R\!\left(\!{\pi\over
 2}\!\right)\!\left|g\right>\nonumber \\
&=&\cos{\theta_1 \over 2}\left|g\right> + i \sin{\theta_1\over 2}\left|e\right>\\
&&\left|\Psi_{\rm final}^{(e)}\right>\nonumber \\
&=&\! T\!\left(\!{\phi \over \delta\omega},\!
\delta\omega\!\right)\!R\!\left(\!\theta_2\!-\!{\pi\over
2}\!\right)\!T\!\left(\!\tau_{\pi\over 2},\!
\delta\omega\!\right)\!R\!\left(\!\theta_1\!-\!\theta_2\!\right)\!T\!\left(\!\tau_{\pi\over
2}, \!\delta\omega \!\right)\!R\!\left(\!{\pi\over 2}\!\right)\!\left|e\right>\nonumber \\
&=&\cos{\theta_2 \over 2}\left|g\right> + i \sin{\theta_2\over 2} e^{i\phi} \left|e\right>
\end {eqnarray}
Therefore, we can generate an arbitrary quantum state from any unknown state via the on-chip quantum feedback.

In this letter, as the first step of such a complete control
of the qubit state,
we demonstrate the fast initialization of a qubit by using our on-chip quantum feedback scheme. 
Initialization is considered an important application for one-qubit quantum feedback control.
There are five criteria for creating a quantum computer, known as the DiVincenzofs Criteria, and the initialization of the qubit is one of them \cite{criteria}.
In our scheme, by choosing $\theta_1=\pi$ and $\theta_2=\pi$, we can initialize a qubit state to an excited state. Our initialization method has an advantage in terms of speed, because all operations including measurement feedback are carried out on the chip, which is a critical difference from the conventional feedback method.

We use a loop with four Josephson junctions to construct a superconducting flux qubit \cite{SCQ4}.
This flux qubit is coupled to a superconducting quantum interference device (SQUID) structure, which is a part of a JBA resonator.
To increase coupling, the qubit and SQUID share a side line.
The JBA resonator was constructed with edge capacitors and a coplanar waveguide. 
At the center of the coplanar waveguide, we design a SQUID and a flux qubit. 
This SQUID works as a nonlinear inductor.
To control the qubit state, we formed a microwave line near the qubit. 
The device was fabricated using the Al angled evaporation method.

All measurements were performed using a dilution refrigerator at a temperature below 50 mK.
Our qubit gap is $\Delta_Q$=3.3GHz and the JBA resonance frequency is 6.5GHz. 
As the operating point approaches the degeneracy point, the coherence time of the qubit increases.
So we set the operating magnetic field near the degeneracy point $(\Delta\Phi=-0.4 {\rm m}\Phi_0, \omega_{\rm qubit}={\rm 3.4 GHz})$.

First, we observed a qubit energy shift during applying a JBA readout pulse. 
By applying a $\pi$ ($2\pi$ pulse), we can prepare a qubit excited (ground) state. 
The frequency of the Ramsey oscillations is equal to the frequency difference between the qubit energy and the applied microwave frequency. 
Thus, we can measure the energy shift by measuring the Ramsey oscillations.
To measure Ramsey oscillations, we apply two ${\pi\over 2}$ pulses during applying the first readout pulse. 
After that, we apply the second readout pulse to measure the final qubit state.
\begin{figure}[h]
\begin{center}
\includegraphics[width=1.0\columnwidth]{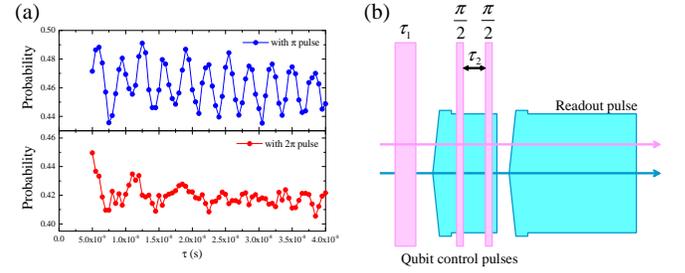}
\caption{
(a) The Ramsey oscillation during an application of a JBA readout pulse.
The blue line shows a Ramsey oscillation when we prepare a qubit excited state by applying a qubit $\pi$ pulse. 
The red line also shows the Ramsey oscillation when we prepare a qubit ground state by applying a $2\pi$ pulse. 
When an excited state of the qubit is detected, the JBA resonator becomes a high amplitude state while the JBA resonator becomes a low amplitude state in case of
 a detection of a qubit ground state.
(b) The pulse sequence for demonstrating a qubit initialization.
Three rectangular pulses are composed of qubit control pulses. 
The first control pulse prepares a qubit initial state. 
The other control pulses are for the initialization. 
The blue two pulses are for JBA readout pulses. 
The first one is for an initialization, and the second one is for detecting a result.
}
\label{ramsey}
\end{center}
\vspace{-0.2cm}
\end{figure}
We show Ramsey oscillations during applying the JBA readout pulse in Fig. \ref{ramsey}(a). 
Here, the oscillation frequencies in the two figures different where we prepare an excited state (upper figure) or a ground state (bottom figure).
The energy shift difference estimated from the oscillation period is around 150MHz.
This result clearly shows that the qubit-energy shift strongly depends on the qubit state detected by the JBA.

Next, we demonstrated a single-qubit quantum feedback scheme.
An applied microwave frequency was tuned to the qubit resonant energy when the JBA detected the ground state of the qubit.
To demonstrate fast qubit initialization, we used the pulse sequence shown in Fig. \ref{ramsey}(b).
We swept both the first pulse width of $\tau_1$, and the pulse distance between ${\pi\over 2}$ pulses as $\tau_2$.
The first pulse prepared an arbitrary superposition of the qubit state. 
The second and the third pulses induced ${\pi\over 2}$ pulses for Ramsey oscillation.

\begin{figure}[h]
\begin{center}
\includegraphics[width=0.8\columnwidth]{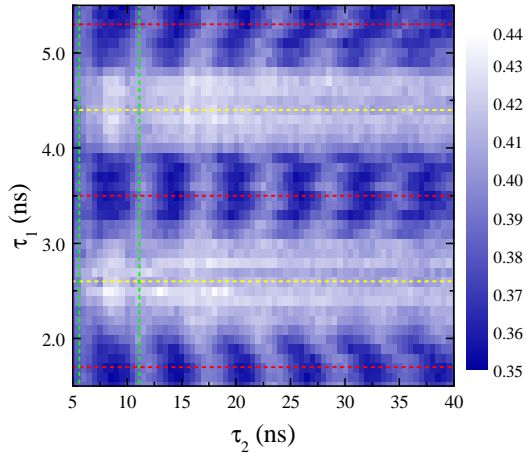}
\caption{
Measurement result of our initialization sequence.
The vertical axis shows the pulse width of the first control pulse $\tau_1$.
The horizontal axis shows the pulse distance $\tau_2$.
The colormap represents the qubit state detected by the second readout pulse where white (dark blue) corresponds to the qubit excited (ground) state Red (Yellow) dashed lines denote $\tau_1$ values when we prepare the qubit excited (ground) state.
Green dashed lines show $\tau_2$ values that the final qubit state does not depend on a prepared qubit state.
The measurement was performed by sweeping $\tau_1$ after sweeping $\tau_2$.
 Due to the time based fluctuations, the Ramsey oscillation period exhibits a small fluctuation.
}
\label{result}
\end{center}
\vspace{-0.2cm}
\end{figure}

We show the measurement results of our fast initialization scheme in Fig. \ref{result}.
When we apply the first qubit control pulse whose length is $\tau_1$= 2.6 ns or 4.4 ns, the qubit is prepared in a ground state.
This means that the JBA state becomes a low amplitude state.
Since the applied microwave frequency is tuned to this qubit energy, Ramsey oscillation does not occur in this case.
On the other hand, when the applied pulse length is $\tau_1$ = 1.7, 3.5, or  5.3 ns, the qubit is prepared in an excited state.
By applying a JBA readout pulse, the JBA state exhibits a high amplitude that shifts the qubit energy, and so we observed a Ramsey oscillation by sweeping $\tau_2$.
Importantly, when we choose $\tau_2$ = 5.5 or 11 ns, the qubit always converges to be an excited state from any type of initial qubit state after performing these operations.
So this result means that we can initialize a qubit state to an excited state.

This initialization sequence is rapid because we do not need to take the readout information from the chip outside the dilution refrigerator or send feedback information from the signal processor to the chip while every previous quantum feedback scheme requires both.
We can roughly estimate the necessary time for this initialization from the sum of the bifurcation time $\tau_{\rm JBA}\equiv Q_{\rm JBA}/f_{\rm JBA}\sim 7$ ns and Ramsey rotation time $\tau_{\pi}\equiv \pi/\delta\omega \sim 5.5$ ns.
So the initialization time is about $10\sim 20$ ns.
This time is as the same as the time required to transmit the signal $2\sim 4$ m in a standard transmission line.
Importantly, this length is shorter than the cable length between a qubit in the refrigerator and the measurement equipment at room temperature.
So our technique demonstrated here has opened new ways to achieve ultra-fast quantum feedback that are not bounded by the length of the transmission line used to connect the chip and the outside of the dilution refrigerator.

In summary, we have succeeded in verifying a fast on-chip quantum feedback by using the selective energy shift of a qubit that depends on the qubit state.
Although our demonstration of on-chip quantum feedback is for a single qubit, our proposal has potential applications to a multi-qubit system.
The information of a measured qubit state can be coded as the microwave amplitude of the JBA output signal, and the JBA output microwave pulse passes through a transmission line.
By coupling the output microwave with the target qubit, the target-qubit energy will also be shifted depending on the microwave amplitude.
So it is possible to perform the selective control of a qubit state depending on the measurement result of another qubit by extending the technique we have described here.
Such a fast on-chip quantum feedback technique is essential for both the realization of quantum information processing and for understanding the foundation of thermodynamics.

This work was supported by the Funding Program for World-Leading Innovative
R\&D on Science and Technology (FIRST Program), JSPS KAKENHI Grant No.
25220601.

\end{document}